\documentclass[pdflatex,sn-nature]{sn-jnl}

\usepackage{graphicx}%
\usepackage{multirow}%
\usepackage{amsmath,amssymb,amsfonts}%
\usepackage{amsthm}%
\usepackage{mathrsfs}%
\usepackage[title]{appendix}%
\usepackage[dvipsnames]{xcolor}%
\usepackage{textcomp}%
\usepackage{manyfoot}%
\usepackage{booktabs}%
\usepackage{algorithm}%
\usepackage{algorithmicx}%
\usepackage{algpseudocode}%
\usepackage{listings}%
\usepackage{bm}
\usepackage{wrapfig}
\usepackage{makecell}
\usepackage{float}
\usepackage{anyfontsize}
\usepackage{hyperref}

\begin{document}

\title[Article Title]{Simulating Pattern Recognition Using Non-volatile Synapses: MRAM, Ferroelectrics and Magnetic Skyrmions}


\author{Luis Sosa}
\author{Minhyeok Wi}
\author{Miguel Barrera}
\author{Imran Nasrullah}
\author{Yingying Wu\footnote{Corresponding author: yingyingwu@ufl.edu}}
\affil{Electrical and Computer Engineering, University of Florida, Gainesville, FL 32611, USA}


\abstract{This project explores the use of non-volatile synapses in neuromorphic computing for pattern recognition tasks through a comprehensive simulation-based approach. The main approach is through spintronic synapses, which leverage the electron's spin properties to achieve efficient data processing and storage. This offers a promising alternative to traditional electronic synapses which require constant power recharge to prevent data leakage. The goal is to develop and simulate a neural network model that incorporates spintronic synapses, examining their potential to perform complex pattern recognition tasks such as image and sound classification. By building a simulation environment, we adopt various models, including spin transfer torque based MRAM, voltage controlled magnetic anisotropy based MRAM, ferroelectric field effect transistors, and skyrmion-based nanotrack for synaptic devices, to evaluate their performance and compare results across different non-volatile implementations. The findings will highlight the effectiveness of spintronic synapses in creating low-power, high-performance neuromorphic hardware, providing valuable insights into their application for future energy-efficient artificial intelligence systems.}




\maketitle

\section{Introduction}
The exponential growth of data-driven applications and artificial neural network have highlighted the limitations of traditional data storage systems, which encode information in binary states, `0' and `1'. It requires constant recharging to maintain the number of electrons to be higher than the threshold, suffering inherent inefficiencies including high energy consumption and significant heat dissipation.
Currently, conventional computers are predominantly built on the von-Neumann architecture, which separates data storage from computation. This architecture creates a fundamental bottleneck as enormous data must be continuously transferred between memory and processing units with the limited size of data bus. The resulting challenges and inefficiencies motivate to increase the demand for energy-efficient, non-volatile, and high-performance computing systems.

Spintronic devices offer groundbreaking solution to these limitations. By leveraging the spin degree of freedom of electrons, spintronics enable non-volatile data storage with high scalability and energy-efficient computation \citep{pan2022efficient,wu2022manipulating,wu2020large, yang2020termination, pan2020observation, wang2020topological}, addressing the bottlenecks inherent to von Neumann architectures. Unlike traditional data storage systems restricted to binary states, spintronic devices can facilitate neuromorphic computing with multiple discrete states, introducing advanced data representation and processing.

Recent advances in spintronic synapses have demonstrated their potential to revolutionize neuromorphic computing by mimicking the behavior of biological synapses with unprecedented energy efficiency. Spintronic synapses leverage physical phenomena to achieve analog-like weight modulation, enabling synaptic plasticity essential for learning processes in artificial neural networks. Recent studies have shown successful implementation of spintronic synapses in both experimental and simulation environments, achieving low-power switching, high endurance, and scalability \citep{marrows_neuromorphic_2024,fukami_perspective_2018,hlone_multilayer_2024}. These developments mark a significant step toward realizing hardware-efficient neuromorphic systems capable of performing complex cognitive tasks with minimal energy consumption.

This work presents a comprehensive comparative analysis of different non-volatile devices for neuromorphic computing and artificial intelligence (AI) applications, evaluating their performance, scalability, and energy efficiency. The focus is on four distinct device architectures: (1) magnetoresistive random access memory (MRAM) with spin-transfer torque (STT), (2) MRAM with voltage-controlled magnetic anisotropy (VCMA), (3) ferroelectric field-effect transistors (FeFETs), and (4) magnetic skyrmion nanotrack devices. NeuroSIM is utilized to demonstrate the integration of these devices into a neural network, highlighting their capability to enable energy-efficient, high-performance AI systems. 

\section{Device Architectures}

\subsection{STT-MRAM}
MRAM is a non-volatile memory device that uses magnetic properties and spin-dependent electron tunneling to store data. The functionality of MRAM is built on the interaction of three primary components: the free layer, the fixed layer, and the oxide layer. The free layer consists of the ferromagnetic layer, and the magnetization direction is dynamically switched between parallel (P) and antiparallel (AP) orientations relative to the fixed layer. In STT-MRAM, the switching is achieved by applying a spin-polarized electric current which transfers angular momentum to the free layer, flipping its magnetization. The magnetic orientation of the free layer toggles between P and AP states based on the polarization of the current, encoding binary data to be stored in the memory cell. The fixed layer has a stable and uniform magnetization direction, which remains unaffected by external influences. It serves as a reference for the magnetization state of the free layer. The oxide layer composed of magnesium oxide (MgO) is located between the free and fixed layers, which facilitates tunneling magnetoresistance (TMR).

The magnetization dynamics of the free layer are governed by the stochastic Landau-Lifshitz-Gilbert (s-LLG) equation, which incorporates external fields, anisotropy, thermal noise, and STT effects \citep{ament_solving_2017}. The equation is expressed as: 

 \begin{equation}
    \frac{d\mathbf{M}}{dt} = -\gamma \mu_\textrm{0} \left(\mathbf{M} \times \mathbf{H}_\textrm{eff} \right) + \frac{\alpha}{M_\textrm{s}} \left( \mathbf{M} \times \frac{d\mathbf{M}}{dt} \right) - \frac{\mathbf{M} \times (\mathbf{M} \times \mathbf{I}_\textrm{s})}{q N_\textrm{s} M_\textrm{s}}
 \end{equation}

where \(\mathbf{M}\) is magnetization vector, \(\mathbf{H}_\textrm{eff}\) is the effective magnetic field, \(\alpha\) is the damping constant, and \(\gamma\) represents the gyromagnetic ratio.

\vspace{-10pt}
\begin{figure}[H]
    \centering
    \includegraphics[width=1\linewidth]{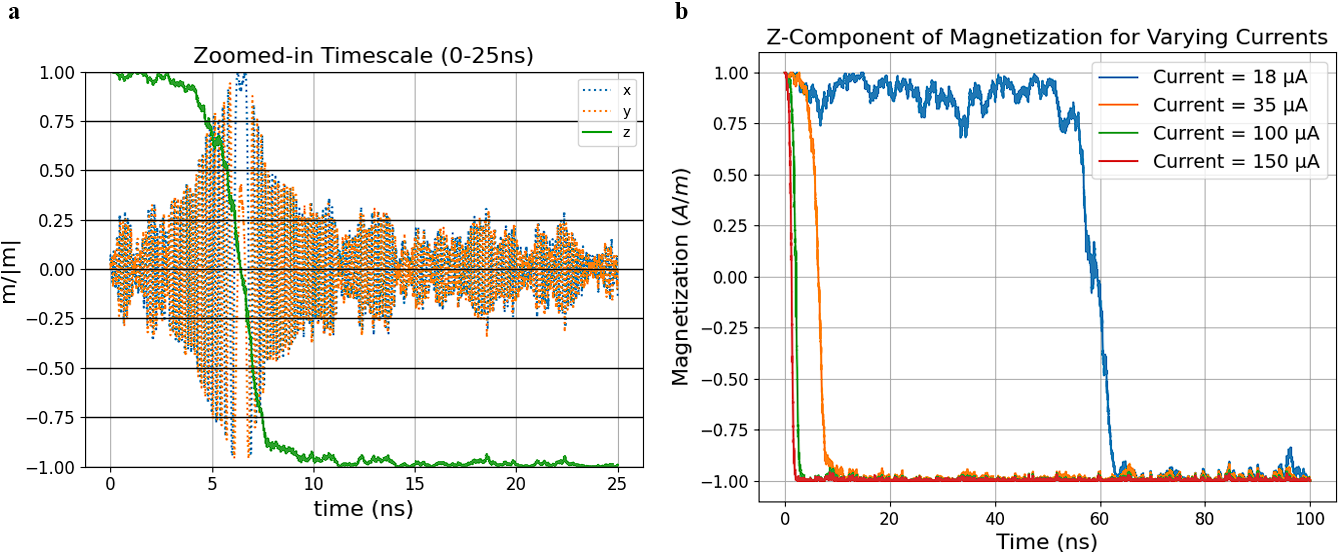}
        \vspace{-15pt}
    \caption{Magnetization switching in STT-MRAM. \textbf{a.} Magnetization vector dynamics in 2D. \textbf{b.} Magnetization vector trajectory for varying current.}
    \label{fig:vect-dynamics}
\end{figure}
    \vspace{-15pt}

Thermal noise plays a significant role in stochastic MRAM operation by acting as a stochastic force that induces oscillations in the magnetization vector of the free layer. The oscillations are inherent to the thermal fluctuations present in the system and can influence the stability of the magnetization state. Thermal noise alone is insufficient to induce the transition between states due to a lack of external torque to drive the switching process. However, as an external torque is applied by spin-polarized current, the magnetization vector is tilted away from the z-axis due to angular momentum on the free layer. As a result, the z-component of the magnetization vector decreases, weakening its alignment with the axis. If the combined effects of thermal noise and applied torque are sufficiently large, the magnetization vector undergoes stochastic switching between the P and AP states. This switching is probabilistic and depends on the magnitude and direction of the applied current, making it possible to control the state transitions through precise current modulation. Fig. \ref{fig:vect-dynamics}a shows how the magnetization vector dynamics change over time with stochastic force induced by thermal noise. 

In Fig. \ref{fig:vect-dynamics}b, it illustrates the z-component of the magnetization vector (\(m_\textrm{z}\)) during switching with varying current magnitudes: 18 $\mu$A, 35 $\mu$A, 100 $\mu$A, and 150 $\mu$A. This z-component reflects the magnetization state of the free layer in MRAM, transitioning between the P state (\(m_\textrm{z} = +1\)) and the AP state (\(m_\textrm{z} = -1\)) due to the applied current. The results demonstrate how current magnitude affects both the feasibility and switching speed. 

The current of 18 $\mu$A represents the lowest threshold required to ensure a successful switch between the P and AP states. At this value, the STT generated by the spin-polarized current is sufficient to overcome the inherent damping forces and induce the required magnetization tilt. Without meeting this threshold, the torque would be insufficient to destabilize the initial state and achieve a full transition. The relationship between current magnitude and switching time is evaluated in the graph. The switching time decreases significantly based on the current values because higher current values induce stronger STT, which accelerates the tilt of the magnetization vector. Conversely, a lower current value results in a slower transition due to the weaker torque. The inverse relationship highlights a trade-off in MRAM design in that higher current magnitudes enable faster switching, but increase power consumption and affect energy efficiency. 

\subsection{VCMA-MRAM}
VCMA, or voltage controlled magnetic anisotropy, is a device in which the easy axis of the magnetization vector $H_\textrm{K}$, the preferred direction of alignment of the tunable magnetization vector within the device, is modulated by the application of a voltage. In this work, a subset of magnetic anisotropy is considered called perpendicular magnetic anisotropy (PMA), in which $H_\textrm{K}$ will be aligned to be perpendicular to the material interfaces that exist within the VCMA device. 

\begin{figure}[hbt!]
    \centering
    \includegraphics[width=0.7\linewidth]{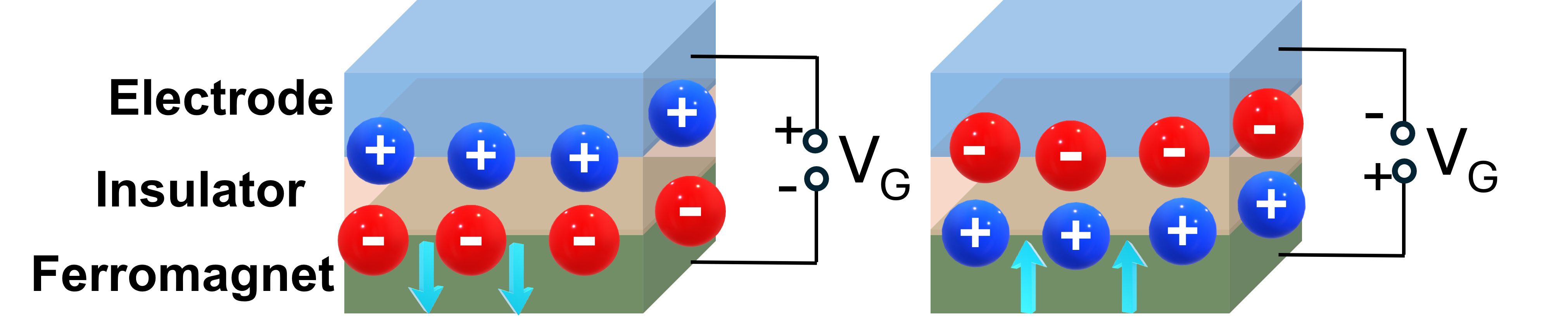}
    \caption{VCMA schematics and switching mechanism. }
    \label{fig:vcma-basic}
\end{figure}

As seen in Fig. \ref{fig:vcma-basic}, typical VCMA devices are constructed via a stacked layers. One example is using Ta/CoFeB/Pt/MgO structure\citep{jeong_spintronic_2024}, in which a voltage $V_\textrm{G}$ is applied across MgO layers. The inclusion of a thin Pt layer at the CoFeB/MgO interface enhances the VCMA effect, enabling the formation and control of a magnetic easy-cone state in CoFeB through the application of the gate voltage.

The following equation shows the dependence of the perpendicular magnetization of the free layer of the VCMA device with an applied voltage:
\begin{equation}
\label{Test}
H_\textrm{Keff}(V) = H_\textrm{K}(V) + H_\textrm{dz} + H_\textrm{ext}
\end{equation}
where $H_\textrm{Keff}$ is the effective magnetization vector along the perpendicular +z-axis, $H_\textrm{K}(V)$ is the voltage-dependent magnetization vector, $H_\textrm{dz}$ is the demagnetization vector, and $H_\textrm{ext}$ is an external magnetization field, typically employed to stabilize the device's magnetics. 

\vspace{-15pt}
\begin{figure}[hbt!]
    \centering
    \includegraphics[width=1\linewidth]{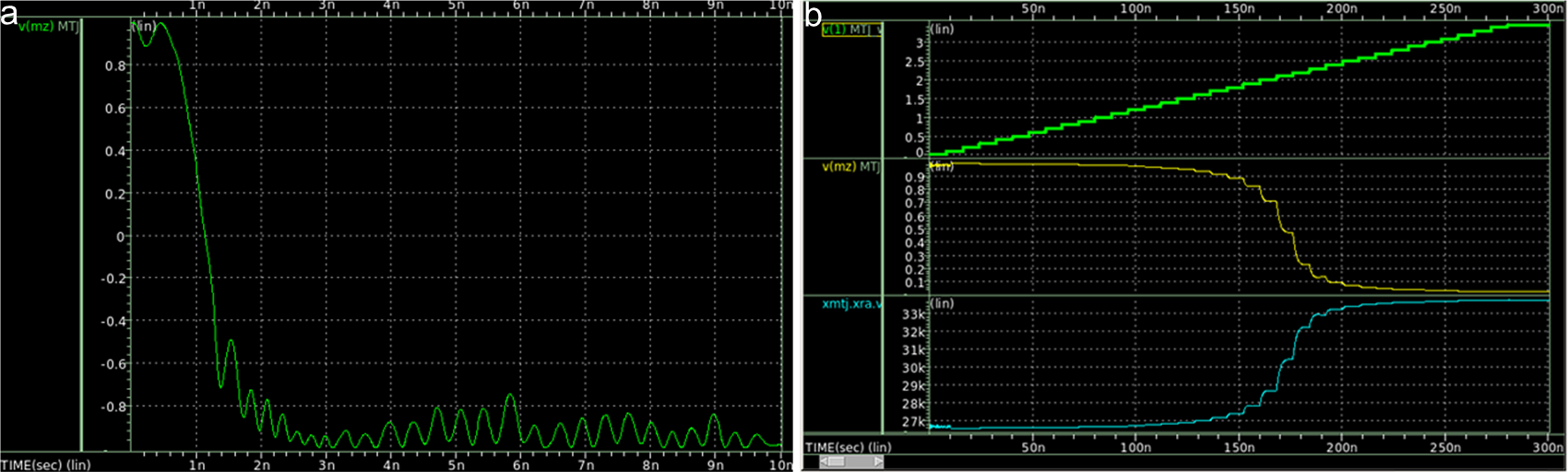}
    \vspace{-15pt}
    \caption{VCMA-MRAM Switching. \textbf{a.} VCMA's $m_\textrm{z}$ switching fully from +1 to -1, displaying capability for analog resistance states. \textbf{b.} VCMA analog states when applying a fine grained voltage step-up pulse.}
    \label{fig:vcma-2}
   \vspace{-15pt}
\end{figure}

Through the application of voltage pulses of varying magnitude to the VCMA device, $H_\textrm{Keff}$ is lowered in magnitude. This causes the magnetic moment within the free-layer $+m_\textrm{z}$ to greatly reduce its preference to align with the +z-axis. From this, a general phenomenon is observed in which $+m_\textrm{z}$ takes quantized values between +1 and -1. Each $m_\textrm{z}$ corresponds to a different resistance level. Therefore, switching within the VCMA device to produce analog states is accomplished by applying voltage pulses to lower the easy axis of the free layer's magnetization.

Finally, to measure the analog resistance of VCMA devices, an ac ${J_\textrm{x}}$ current density is applied through drain and source of the device corresponding to current-conducting layers that are not the free layer. From here, the Hall resistance $R_\textrm{H}$ is measured. Since $R_\textrm{H}$ is a function of the perpendicular magnetization applied to it, its magnitude will change based on the varying magnitudes of $m_\textrm{z}$.
An extremely small (nA-pA), non-polarized current is vertically going through VCMA device stack. Different $m_\textrm{z}$ will lead to different resistance values, corresponding to different analog states.

To simulate the VCMA device, a simplified device stack is used, as well as a diagram showing the magnetic change that is produced within this device stack upon application of the VCMA voltage:
Application of the VCMA voltage to the free and oxide layer, similar to Fig. \ref{fig:vcma-basic}, causes the magnetization of $m_\textrm{z}$ to change to produce different magnetization states. From here, several parameters of the VCMA model were modified to encourage varying analog states of the VCMA according to magnetic equations that describe the VCMA device's dynamically changing magnetic vector:

\begin{enumerate}
    \item $H_\textrm{ext}$ was changed from 20 A/m to -10 A/m. This lessened the stability provided from an external magnetic field, which``encourages" the $m_\textrm{z}$ value to reduce its preference for +1. This makes it likely to take different quantized values between +1 and -1.
    \item VCMA coefficient $\epsilon$ was reduced from 7.6$\times 10^{-14}$ $JV^{-1}m^{-1}$ to 4$\times 10^{-14}$ $JV^{-1}m^{-1}$ . With a lower value of $\epsilon$, the VCMA voltage is less likely to take a drastic change from any of the intermediate $m_\textrm{z}$ and ``floor" to -1.
    \item ${t_\textrm{ox}}$, the free layer's thickness, was changed from 1.6 nm to 1.0 nm. Reducing the thickness of the free layer reduces the strength of the magnetization that is possible in the VCMA device when its $m_\textrm{z}$ is at +1. Thus, the magnetization of the free layer is now less inclined to ``ceil" up +1.
\end{enumerate}

With these changes in mind to create analog states, a varying VCMA pulse was applied from 0 V to +3.45 V with steps of +0.1 V, as shown in Fig. \ref{fig:vcma-2}b. With each increase in the VCMA voltage, the free layer's $m_\textrm{z}$ value drops incrementally. This change in its resistance states produces an analog functionality, allowing for a device architecture that supports multiple states.

\subsection{FeFET}

\begin{figure}[H]
\vspace{-15pt}
    \centering
    \includegraphics[width=1\linewidth]{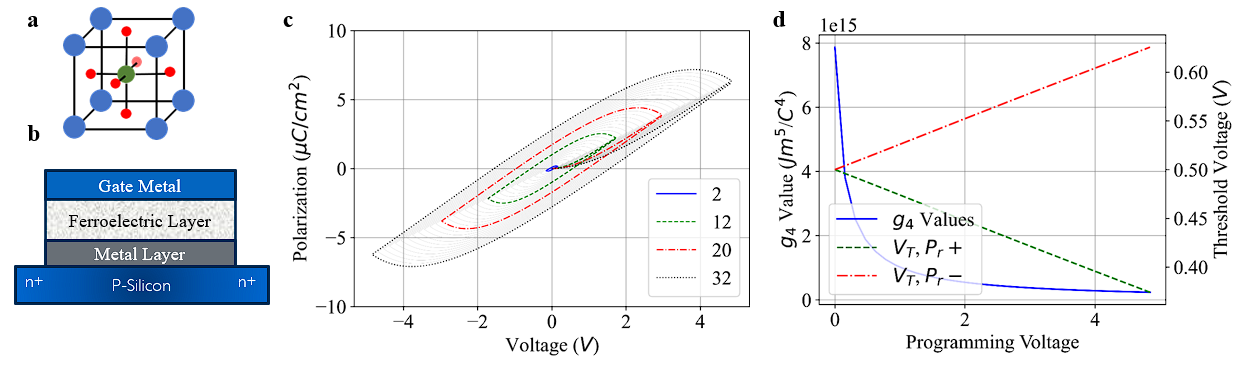}
    \vspace{-15pt}
    \caption{FeFET schematics and simulation results. \textbf{a.} Symmetric configuration of $BaTiO_\textrm{3}$. \textbf{b.} Metal-ferroelectric insulator-metal Architecture. \textbf{c.} Hysteresis loops for FeFET. \textbf{d.} Relationship between LK parameter, $V_\textrm{T}$, and programming voltage.}
    \label{fig:fefet}
    \vspace{-15pt}
\end{figure}

FeFETs leverage the spontaneous polarization of ferroelectric materials to enable non-volatility and multi-bit data storage.
The core of FeFET functionality lies in the ferroelectric crystal structure observed in Fig. \ref{fig:fefet}a, where ions within the lattice shift under an electric field to create spontaneous polarization. By integrating a ferroelectric layer (Fig. \ref{fig:fefet}b) into traditional CMOS structures, FeFETs enhance functionality with minimal changes to existing fabrication processes. For example, in materials like hafnium oxide, the displacement of central ions from their symmetric positions generates an electric dipole, breaking inversion symmetry \citep{zagni_reliability_2023}. This behavior gives rise to the characteristic hysteresis loop, where polarization states can be switched and retained even after the electric field is removed. These stable polarization states modulate the transistor's threshold voltage $V_\textrm{T}$, enabling FeFETs to store multiple discrete states and mimic the analog functionality of biological synapses. 

Polarization dynamics are modeled using the Landau-Khalatnikov equation \citep{aziz_physics-based_2016}, which characterizes the time-dependent response of polarization under an applied electric field. It can be described as 
\begin{equation}
\label{lankha}
E - \rho \frac{dP}{dt} = a P + b P^3 + c P^5
\end{equation}
where $a$, $b$, and $c$ are thermodynamic constants that define the material's behavior; $a$ determines polarization stiffness and depends on the temperature relative to the Curie point, $b$ stabilizes the polarization by introducing higher-order terms that govern energy well depths, while $c$ ensures further stability by preventing divergence. The coefficient $\rho$ reflects the system's viscosity, controlling the speed at which polarization responds to an electric field \citep{maslovskaya_theoretical_2021}. 
To align with device-specific operational characteristics, we mapped the polarization states to precisely 32 discrete conductance levels. This approach balances the precision required for analog neuromorphic computing with practical constraints, as derived from literature\citep{jerry_ferroelectric_2017}. The states correspond to incremental shifts in the ferroelectric layer's polarization, modulating the transistor's threshold voltage. 

FeFET-based synapses mimic biological synaptic plasticity by modulating their conductance states to represent synaptic weights. This is achieved by applying voltage pulses to update the polarization state incrementally \citep{jerry_ferroelectric_2017}. The amplitude, duration, and polarity of the voltage pulses are adjusted to control the synapse's potentiation and depression. The resulting conductance changes are stored in a non-volalitely, enabling long-term weight retention without continuous power supply.

In our FeFET modeling, the generated hysteresis loops for devices configured with 2, 12, 20, and 32 states (shown in Fig. \ref{fig:fefet}c) illustrate the polarization dynamics under varying programming voltages. These loops highlight how increasing the number of discrete states enhances the analog precision of FeFET-based synapses, enabling finer granularity in weight representation. Additionally, the influence of the Landau-Khalatnikov parameter $g_\textrm{4}$ on the polarization response was studied. Fig. \ref{fig:fefet}d includes the threshold voltage $V_\textrm{T}$ as a function of programming voltage, providing insights into the direct relationship between device parameters and programming inputs. 

\subsection{Magnetic Skyrmions}

\begin{figure}[H]
\vspace{-15pt}
    \centering
    \includegraphics[width=1\linewidth]{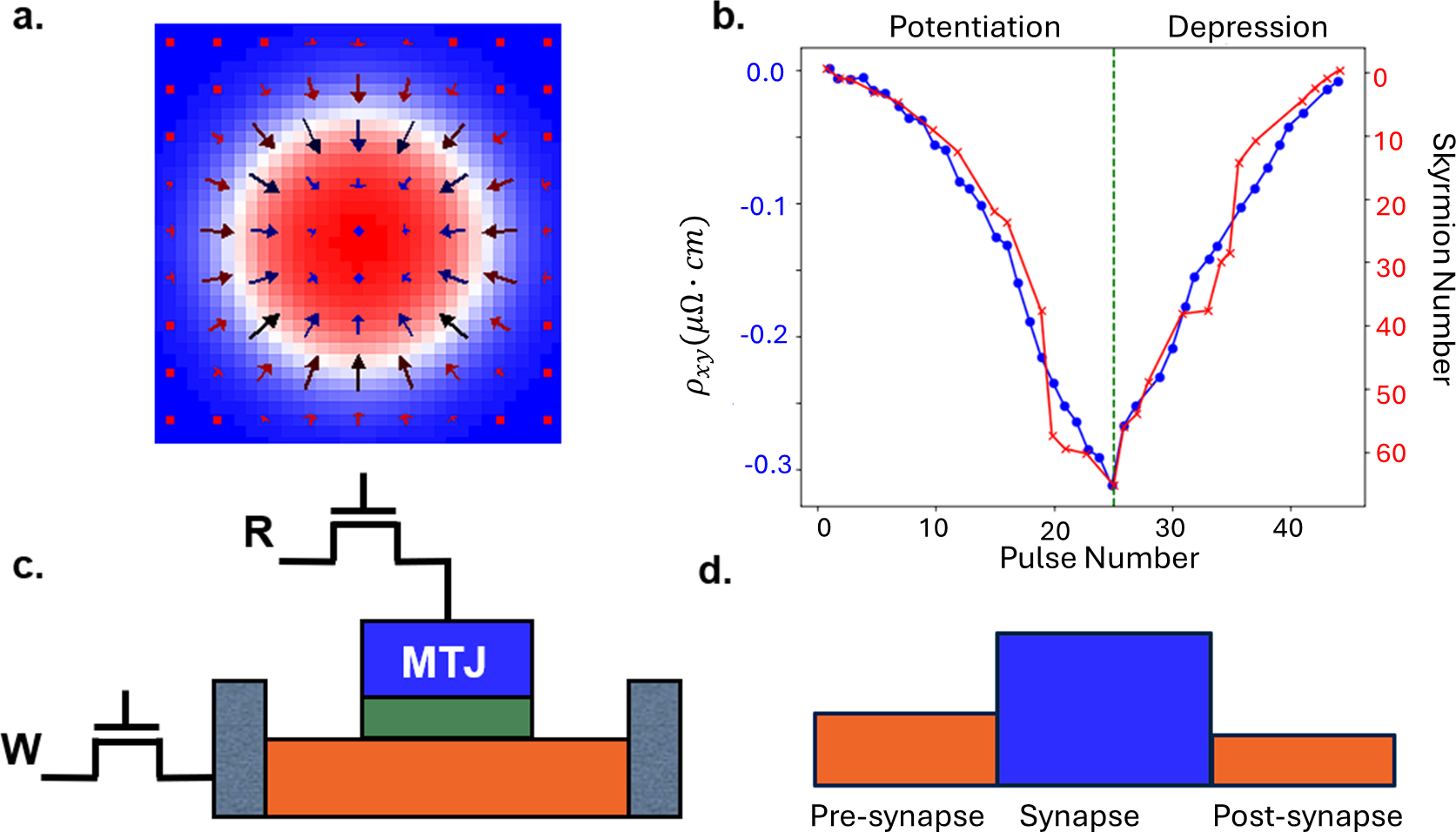}
        \vspace{-15pt}
    \caption{Artificial synapse based on magnetic skyrmions. \textbf{a.} Simulation of N\'eel-type skyrmion in ferromagnetic material. The background color represents the magnetization in the $z$-axis, with red showing magnetization into the page and blue showing out of the page. \textbf{b.} Schematics of measured Hall resistivity change and number of skyrmions as a function of injected pulse number. \textbf{c.} Top view schematic of skyrmion synapse device in a 2T1R architecture. \textbf{d.} Side view schematic of skyrmion synapse device in a 2T1R architecture.}
    \label{fig:skyrmion}
    \vspace{-15pt}
\end{figure}

Magnetic skyrmions are nanoscale, topologically stable spin textures with particle-like behavior \citep{wu2024magnetic,zhang20242d,huang_magnetic_2017,zhong2024integrating,wu2022van, wu2020neel,hu_magnetic_2023, fert_magnetic_2017}. Micro-magnetic simulation of single magnetic skyrmion (Fig. \ref{fig:skyrmion}a) was performed with OOMMF software \citep{donahue_m_j_oommf_1999}. They exhibit unique physical and topological properties, such as defined chirality and quantized topological charge, making them promising candidates for various technological applications, particularly in spintronics and neuromorphic computing \citep{song_skyrmion-based_2020,vakili_skyrmionicscomputing_2021,li_magnetic_2017}.

Skyrmions form due to the interplay of chiral interactions, primarily the Dzyaloshinskii-Moriya interaction (DMI) \citep{dzyaloshinsky_thermodynamic_1958,moriya_anisotropic_1960}, which takes the form of:

\begin{equation}
\label{dmi}
H_\textrm{DMI} = \bm{d}_\textrm{1,2} \cdot (\bm{S}_\textrm{1} \times \bm{S}_\textrm{2})
\end{equation}

where $\bm{d}_\textrm{1,2}$ is the Dzyaloshinskii-Moriya vector and $\bm{S}_\textrm{1}$ and $\bm{S}_\textrm{2}$ are the neighboring spins. The DMI is an asymmetric exchange interaction from strong spin-orbit coupling, which produces an orthogonal alignment of neighboring spins with chirality determined by the direction of the rotation from $\bm{S}_\textrm{1}$ to $\bm{S}_\textrm{2}$ around $\bm{d}_\textrm{1,2}$. Two common types of skyrmions are N\'eel-type skyrmions, found in systems with interfacial DMI, and Bloch-type skyrmions, associated with bulk DMI. Their spin configurations vary, but both feature a continuous and topologically protected rotation of magnetization. The topological charge, or skyrmion number, quantifies the winding of the magnetization, distinguishing skyrmions from other magnetic textures like magnetic bubbles \citep{han_skyrmions_2017}.

These structures were first observed in non-centrosymmetric bulk materials such as MnSi \citep{muhlbauer_skyrmion_2009} and later in multilayered thin films like Pt/CoFeB/MgO \citep{woo_observation_2016}. Thin films, especially those with PMA, are particularly favorable for practical applications due to their stability at room temperature and compatibility with device fabrication.

Skyrmion dynamics are governed by their solitonic nature, allowing them to behave as quasi-particles. They can be created, annihilated, and manipulated using electrical currents. STT and SOT are key mechanisms for their motion \citep{fert_magnetic_2017}, where the skyrmion's gyrotropic motion results in the skyrmion Hall effect: a deviation from the direction of the applied force. Skyrmions are resilient against external perturbations like defects, thanks to their topological protection. However, defects and boundary interactions can still influence their stability and motion \citep{je_direct_2020}, with high currents potentially driving skyrmions out of a system.

Fig. \ref{fig:skyrmion}b shows the linear relationship observed between the number of skyrmions, the change in resistivity and the number of current pulses during the potentiation and depression stages\cite{song_skyrmion-based_2020}. Using these distinct conductance states, we can generate up to 32 different synaptic weights using one skyrmion memory cell, which would take 5 cells with traditional memory architectures. We used an existing synaptic device architecture \citep{song_skyrmion-based_2020} using a 2T1R cell structure (Fig. \ref{fig:skyrmion}c). In this configuration, the potentiation process begins by injecting a series of square wave current pulses of amplitude $|j_\textrm{a}| = 4 \times 10^{10} A\cdot m^{-2}$ and duration $t_\textrm{p} = 0.05$ ns along the magnetic material, allowing the use of current-induced SOTs to control the nucleation of N\'eel-type skyrmions in the presynapse area (Fig. \ref{fig:skyrmion}d), which are then driven towards the active synapse area. Once there, the out-of-plane magnetization resulting from the skyrmions shows a linear dependence to the Hall resistivity, so the resistance of the device will be directly proportional to the number of skyrmions in the synaptic area. This phenomenon is exploited to create different conductance levels depending on the number of skyrmions nucleated, which depends on the number of current pulses. The conductance of the device is then read through the MTJ to create the synaptic weights. The depression process occurs by inverting the polarity of the current pulses while maintaining their amplitude and duration to excite the skyrmion texture, leading to skyrmion annihilation.

\subsection{Device Comparison}

The comparison of neuromorphic devices reveals distinct trade-offs in terms of physical footprint, speed, and functionality. Table \ref{table:device-comp} provides a detailed summary of the key parameters for skyrmion-based memory, FeFETs, STT-MRAM, and VCMA-MRAM, highlighting their strengths in blue colors.

\begin{table}[h]
\vspace{-15pt}
    \centering
    \caption{Summary of device parameters among four types of neuromorphic devices, with best parameters highlighted.} 
    \normalsize
    \begin{tabular}{|c|c|c|c|c|}
    \hline
        Parameter & Skyrmion \citep{song_skyrmion-based_2020} & FeFET & STT & VCMA \\ \hline \hline
        Cell Area & 616 $F^2$ & 303 $F^2$ & \textcolor{blue}{48 $F^2$} & \textcolor{blue}{48 $F^2$} \\ \hline
        Synapse Area & 616 $F^2$ & 303 $F^2$ & 288 $F^2$ & \textcolor{blue}{48 $F^2$} \\ \hline
        Read Speed &\textcolor{blue}{ 1 ns} & 75 ns & 5.75 ns & 1 ns \\ \hline
        Write Speed & \textcolor{blue}{0.05 ns} & 75 ns & 5.75 ns & 1 ns \\ \hline
        On/Off Ratio & 3 & \textcolor{blue}{45} & 3.5 & 1.5 \\ \hline
        \# of states &\textcolor{blue}{32} & \textcolor{blue}{32} & 2 & 6 \\ \hline
    \end{tabular}
    \label{table:device-comp}
 \vspace{-15pt}
\end{table}

To evaluate their physical footprint, we use the concepts of cell area, defined as the area of one memory device, and synapse area, defined as the area required to store one synaptic weight. In most cases, we used a synapse area corresponding to a 5-bit synapse, except for the VCMA devices where the highest number of stable states obtained was 6. Among the evaluated devices, VCMA-MRAM exhibits the smallest cell and synapse area, benefiting from its minimalistic design optimized for integration density. STT-MRAM, although similar to VCMA-MRAM at the cell level, demonstrates a significantly larger synapse area due to its binary operation. FeFETs, in contrast, feature equal cell and synapse areas, which are relatively larger due to the need for ferroelectric layers and CMOS compatibility. Skyrmion-based synapses have the largest cell and synapse areas due to their reliance on complex magnetic domains and multilayer thin-film architectures.

The fastest write speeds are achieved by skyrmion-based memory and VCMA-MRAM, both operating at 0.05 ns, with read speeds of 1 ns. FeFETs, while slower, offer robustness through a high on/off ratio that provides stability against state changes during read and write operations. Conversely, VCMA-MRAM’s low on/off ratio renders it more sensitive to perturbations and noise, which may impact reliability in practical applications.

STT-MRAM is limited to binary state storage, restricting its applicability in analog and multi-level neuromorphic systems. In contrast, FeFETs, VCMA-MRAM and skyrmion-based memory can support up to 32 discrete states, offering significant advantages for analog neuromorphic computing and multi-level storage. This capability allows for finer granularity in weight representation, essential for tasks requiring high precision.

\section{Neuromorphic Pattern Recognition}
We utilized NeuroSim, a robust hierarchical simulation framework, to model and evaluate FeFET-based memory arrays for neuromorphic applications. This framework operates across multiple levels of abstraction, integrating device-level characteristics, circuit-level array configurations, and algorithmic performance. NeuroSim is instrumental in simulating and analyzing emerging non-volatile memory devices, offering precise insights into trade-offs between digital and analog memory technologies in neuromorphic computing compared to traditional CMOS \citep{peng_dnnneurosim_2019}\citep{peng_dnnneurosim_2021}.

\subsection{Memory Array}
At the chip level, the architecture is organized into tiles, which are interconnected through a hierarchical structure such as an H-tree. Each tile serves as an independent computational unit, comprising buffers, processing elements (PEs), and synaptic arrays. These tiles enable high parallelism, which is critical for mimicking the massive interconnected networks of biological neural systems \citep{peng_dnnneurosim_2019}. 

Inside each processing element, synaptic arrays form the core of computation and can operate in either digital or analog modes. For digital operation, the eNVM are implemented as part of 1T1R or 2T1R configurations, where binary states are stored and processed row by row \citep{peng_dnnneurosim_2021}. This setup uses decoders, sense amplifiers, and multiplexers to execute precise read and write operations while maintaining area efficiency. Weighted sums are computed iteratively by cycling through binary inputs and accumulating results, providing high precision.

For analog operation, crossbar arrays leverage their inherent multiconductance states to directly store analog weights. Row drivers apply voltage inputs, and column integrators aggregate currents to compute weighted sums. Analog-to-digital converters (ADCs) convert these currents into digital signals for further processing \citep{peng_dnnneurosim_2021}. To adjust weights during training, voltage pulses are used to modify the polarization states of analog eNVM incrementally, enabling highly precise weight updates. The analog mode's parallel computation significantly reduces latency for matrix-vector multiplications compared to digital row-by-row processing.

\subsection{Neural Network}

For pattern recognition, we used a modified version of the VGG8 network; a compact and efficient convolutional neural network (CNN) derived from the popular VGG (Visual Geometry Group) architecture family \citep{simonyan_very_2015}.  
The network alternates convolutional layers with ReLU activations and max-pooling layers, followed by fully connected layers toward the output. The simplicity of VGG8 ensures ease of implementation and faster training times compared to deeper architectures, while still achieving competitive accuracy for various image classification tasks. This architecture includes the use of small 3$\times$3 convolutional filters, which are stacked to capture complex features while maintaining computational simplicity. The network architecture is summarized in Table \ref{table:vgg8}.

\begin{table}[hbt!]
\vspace{-5pt}
    \centering
        \caption{Neural network parameters.} 
    \normalsize
    \begin{tabular}{|c||c|c|c|c|}
    \hline
        Layer & Type & Structure & Kernel & Stride \\ \hline  \hline
        1 & Convolutional & 32$\times$32$\times$3 & 3$\times$3 & 1$\times$1 \\ \hline
        2 & Convolutional & 32$\times$32$\times$128 & 3$\times$3 & 1$\times$1 \\ \hline \hline
        \multicolumn{5}{|c|}{maxpool} \\ \hline \hline
        3 & Convolutional & 16$\times$16$\times$128 & 3$\times$3 & 1$\times$1 \\ \hline
        4 & Convolutional & 16$\times$16$\times$256 & 3$\times$3 & 1$\times$1 \\ \hline\hline
        \multicolumn{5}{|c|}{maxpool} \\ \hline \hline
        5 & Convolutional & 8$\times$8$\times$256 & 3$\times$3 & 1$\times$1 \\ \hline
        6 & Convolutional & 8$\times$8$\times$512 & 3$\times$3 & 1$\times$1 \\ \hline \hline
        \multicolumn{5}{|c|}{maxpool} \\ \hline \hline
        7 & Fully Connected & 8192 & 1$\times$1 & 1$\times$1 \\ \hline
        8 & Fully Connected & 1024 & 1$\times$1 & 1$\times$1 \\ \hline 
    \end{tabular}
    \label{table:vgg8}
    \vspace{-15pt}
\end{table}

The reduced depth and parameter count make VGG8 well-suited for applications requiring a balance between performance and efficiency. Its reduced model size also makes it more memory-efficient, enabling deployment in resource-constrained environments. Despite being a simplified version, VGG8 demonstrates the enduring versatility of the VGG design principles, proving effective for our purposes.

This CNN is trained on the CIFAR-10 dataset \citep{krizhevsky_learning_nodate}, a widely used benchmark dataset in machine learning, particularly for image classification tasks. It consists of 60,000 32$\times$32 color images evenly divided into 10 distinct classes, such as airplanes, cars, cats, and dogs. Each class contains 6,000 images, with 50,000 images for training and 10,000 for testing.

In this study, we trained the model for 100 epochs, storing the weights and gradients in the memory arrays built from the devices of interest. To achieve best results, we implemented a learning rate reduction strategy, where the model's learning rate is divided by 8 at 40 and 80 epochs, allowing it to fine tune its learning process and helping it reach a local minimum of error.

\section{NeuroSim simulation and results}

The results of the NeuroSim simulations, summarized in Table \ref{table:neurosim-results}, provide a comparative evaluation of emerging spintronic neuromorphic devices against conventional SRAM technology, which serves as a baseline. As a mature and widely utilized memory technology, SRAM is renowned for its high-speed operation and energy efficiency, albeit with the trade-off of large area consumption. Across most evaluated devices, including SRAM, the neural network achieved over 90\% accuracy, highlighting that this upper accuracy limit is more dependent on the network architecture than on the specific device characteristics. However, the VCMA device has a lower accuracy of 80\%. This discrepancy is likely due to the limited synaptic weight resolution arising from its small number of states. These findings suggest that there exists a critical threshold in the number of states required for optimal accuracy, beyond which additional states provide diminishing returns.

\begin{table}[hbt!]
\vspace{-15pt}
    \centering
     \caption{Summary of NeuroSim simulation results for all evaluated memory devices.} 
     \small
    \begin{tabular}{|l||l|l|l|l|l|}
    \hline
        Metric & SRAM & Skyrmion & FeFET & STT-MRAM & VCMA-MRAM \\ \hline \hline
        Accuracy & \textcolor{blue}{91.02\%} & \textcolor{blue}{91.18\%} & \textcolor{blue}{90.62\%} & \textcolor{blue}{90.77\%} & 80.27\%\\ \hline
        TOPS & 0.873 & 1.279 & 1.092 & 1.117 & \textcolor{blue}{1.363} \\ \hline
        TOPS/W &1.699 & 2.134 & \textcolor{blue}{2.001} & 2.05 & 2.022 \\ \hline
        Total Latency (s) & 210  & 144  & 364  & 165  & \textcolor{blue}{135 } \\ \hline
        Total Energy (J) & 108 & 86  & \textcolor{blue}{84 } & 89  & 91  \\ \hline
        Leakage Power ($\mu$W) & 2297 & \textcolor{blue}{184 } & 1810  & 1740  & 205  \\ \hline
    \end{tabular}
    \label{table:neurosim-results}
    \vspace{-15pt}
\end{table}

The performance of each device was further evaluated in terms of tera operations per second (TOPS) and tera operations per second per watt (TOPS/W). In terms of TOPS, SRAM demonstrated the weakest performance, while the VCMA device emerged as the top performer, closely followed by the skyrmion-based device. This advantage is attributable to the inherently fast read and write speeds of spintronic and skyrmion-based technologies.
When considering TOPS/W as a measure of energy efficiency, FeFET outperformed all other technologies due to its exceptionally low energy consumption. This result highlights FeFET’s potential for energy-constrained applications, although this efficiency comes with trade-offs in other performance metrics.

Latency was another key factor in the comparative analysis. Despite its high read and write speeds, FeFET performed poorly in latency tests, lagging behind the other devices. By contrast, the VCMA, STT, and skyrmion devices demonstrated superior latency performance, surpassing that of SRAM and making them more suitable for high-speed applications.
In terms of total energy consumption, SRAM exhibited the highest energy usage and leakage power, emphasizing its inefficiency in these areas. FeFET achieved the lowest energy consumption, closely followed by skyrmion devices, which also recorded the lowest leakage power among all technologies. These results position skyrmion-based devices as highly promising candidates for energy-efficient and low-leakage neuromorphic systems, though some limitations remain to be addressed.
Finally, the sensitivity of these devices to noise was evaluated by introducing Gaussian noise with a standard deviation of 500 $\mu$V. This setup emulates real-world noise sources, such as power supply fluctuations, voltage droops, and crosstalk in interconnect lines.

\begin{figure}
    \centering
    \includegraphics[width=0.9\linewidth]{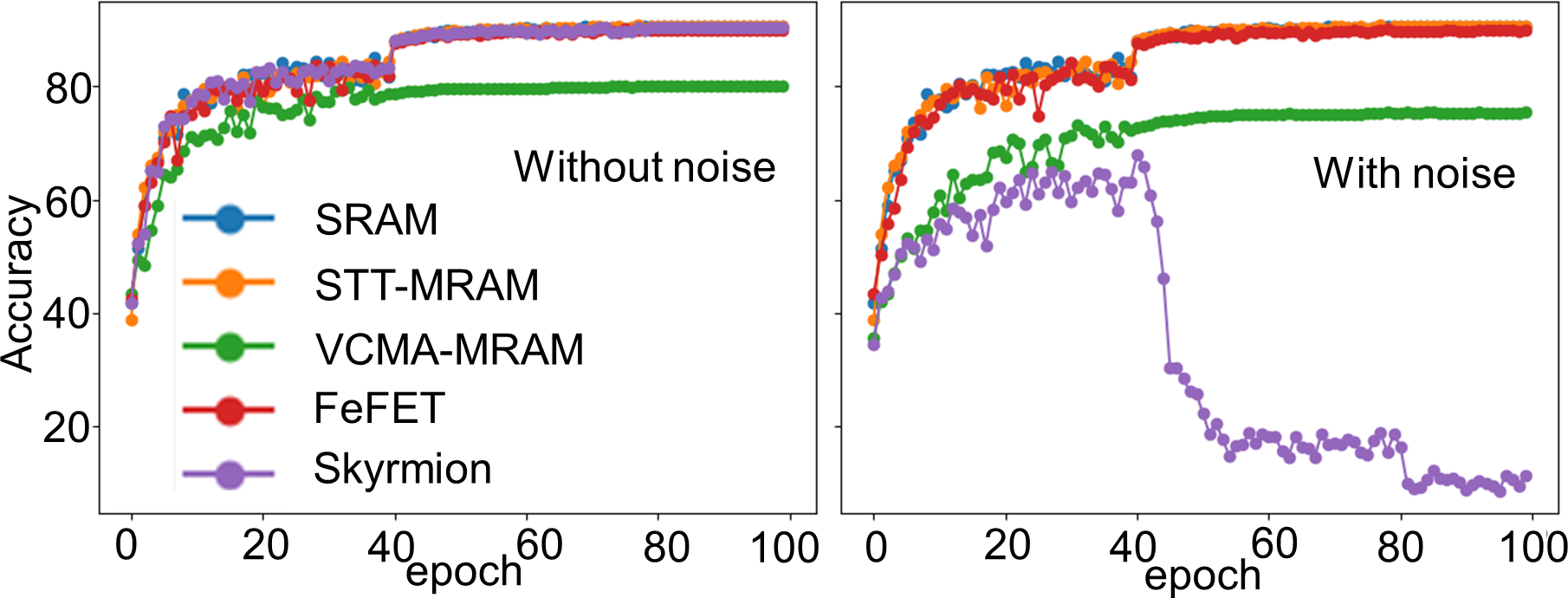}
    \caption{Accuracy results of NeuroSim simulations with and without artificially-added Gaussian noise.}
    \label{fig:noise}
    \vspace{-10pt}
\end{figure}

The data presented in Fig. \ref{fig:noise} shows how most devices exhibited robustness to low levels of noise, with notable exceptions being the VCMA and skyrmion devices. The VCMA device experienced a modest accuracy reduction of approximately 5\%, while the skyrmion device demonstrated a much more pronounced sensitivity, with its accuracy plummeting to 67.5\% under noisy conditions. This degradation was particularly evident at epochs 40 and 80, corresponding to learning rate reductions. As the learning rate decreased, the weight gradient became insufficient to counteract the noise, allowing it to dominate the system's behavior.

The root cause of the skyrmion device's vulnerability lies in its small on/off ratio of 3 combined with a relatively high number of states of 32. These factors result in narrow conductance intervals between adjacent states, making the device susceptible to noise-induced state misclassification. For comparison, the VCMA device, with an even smaller on/off ratio of 1.5, is less affected due to its lower number of states. Conversely, FeFET, which shares the same number of states as the skyrmion device, boasts an on/off ratio 15 times larger, enabling it to resist noise more effectively. Similarly, the STT-MRAM device mitigates noise sensitivity by offering binary states despite its small on/off ratio.

This analysis underscores that noise sensitivity in neuromorphic devices is intricately tied to both the on/off ratio and the number of states. For skyrmion-based devices, achieving an optimal balance between these parameters is critical. By improving the on/off ratio or reducing the number of states, it may be possible to enhance the robustness of skyrmion devices under realistic operating conditions.

\section{Conclusion}
This study provides a comprehensive analysis of spintronic devices for neuromorphic computing, with a focus on their potential for AI applications. The comparison of various technologies, including VCMA-MRAM, STT-MRAM, FeFETs, and skyrmion-based memory, highlights their distinct strengths and weaknesses in terms of performance, scalability, energy efficiency, and noise sensitivity.

The results demonstrate that while VCMA-MRAM excels in speed and energy efficiency, its lower number of states limits its accuracy, especially for high-precision tasks. In contrast, FeFETs, despite their slower write speeds and higher latency, emerge as the most energy-efficient option, making them ideal for applications where power consumption is a critical factor. Skyrmion-based devices, with their fast speeds and large state capacity, offer promising potential for neuromorphic computing but require further optimization to address noise sensitivity and improve robustness. STT-MRAM, although less versatile in terms of state storage, maintains strong performance in terms of reliability and speed.

NeuroSim simulations underscore the importance of device-level characteristics in neural network accuracy, highlighting that the number of states plays a significant role in achieving optimal performance. However, it is evident that performance metrics such as energy efficiency, speed, and latency are highly dependent on the specific application, with trade-offs existing between them.

In conclusion, the ongoing development of spintronic devices holds great promise for advancing neuromorphic computing and AI systems, with each device offering unique advantages suited to different computational requirements. Future work should focus on refining device architectures, enhancing robustness against noise, and optimizing the balance between performance and energy efficiency to enable the next generation of intelligent, energy-efficient systems.

\backmatter

\subsection*{Acknowledgment}
Support from UF Gatorade award and Research Opportunity Seed Fund are kindly acknowledged. This material is also based upon work supported by the National Science Foundation.

\subsection*{Declaration of competing interest}
The authors declare that they have no known competing financial interests or personal relationships that could have appeared to influence the work reported in this work.

\subsection*{Data Availability}
The datasets generated during and/or analysed during the current study are available in the \href{https://github.com/sosapio/Spintronic-Pattern-Recognition}{GitHub repository}.

\subsection*{Author Contribution}
Y. Wu conceived the ideas and supervised the project. L. Sosa led the manuscript writing and discussions of the results, with contributions from M. Wi, M. Barrera, and I. Nasrullah. L. Sosa, M. Wi, M. Barrera, and I. Nasrullah wrote and implemented the simulation, with L. Sosa focusing on skyrmions, M. Wi on STT-MRAM, M. Barrera on FeFET, and I. Nasrullah on VCMA-MRAM. L. Sosa synthesized the required parameters and implemented the neural network simulations in NeuroSim.

\newpage
\bibliography{Manuscript}

\end{document}